\documentclass[epj]{svjour}
\usepackage{graphics}
\begin{document}
\title{Strange prospects for LHC energies}
\author{B.~Hippolyte
\thanks{\emph{Present address: } hippolyt@in2p3.fr}%
~for the ALICE Collaboration
\thanks{A list of all members of this Collaboration is given at the end of this issue.}%
}                     
%
%
\institute{Institut Pluridisciplinaire Hubert Curien, D\'{e}partement de Recherches Subatomiques and Universit\'{e} Louis Pasteur, Strasbourg, FRANCE.}
\date{Received: date / Revised version: date}
%
\abstract{
Strange quark and hadron production will be studied at the
Large Hadron Collider (LHC) energies in order to explore the
properties of both pp and heavy-ion collisions.
The ALICE experiment will be specifically efficient in the
strange sector with the identification of baryons and mesons
over a wide range of transverse momentum.
Dedicated measurements are proposed for investigating
chemical equilibration and bulk properties.
Strange particles can also help to probe kinematical regions
where hard processes and pQCD dominate.
We try to anticipate here several ALICE analyses to be
performed as the first Pb--Pb and pp data will be available.
\PACS{
      {25.75.-q}{describing text of that key}   \and
      {25.75.Dw}{describing text of that key}
     } 
} 
\maketitle
\section{Introduction}
\label{intro_bohq06}
%
%
The ALICE experiment at LHC will be a unique tool for
understanding strange quark and hadron productions~\cite{Carminati:2004fp}.
In relativistic heavy ion collisions, strange particles are
used to probe the chemical equilibration of the cooling Quark
Gluon Plasma (QGP) which can be created in such high energy
density conditions.
The extraction of the transverse momentum ($p_{T}$) spectra
for identified particles is interesting for distinguishing
between different hadronization scenarios in the intermediate
$p_{T}$ range.
In elementary hadronic collisions, strange particles can help
characterizing the underlying event structure and defining a
reference for baryon creation.
We present here some of the measurements which will be
performed with the ALICE experiment both in Pb--Pb and pp
collisions.
%

%
%
In the second section, we discuss the chemical composition
analysis resulting from particle spectra and yields in Pb--Pb
collisions.
Particle identification (PID) capabilities and $p_{T}$ range
for hyperons in ALICE are then summarized.
The third section is dedicated to strange baryon/meson ratio
as a function of $p_{T}$, starting with the importance of
this measurement with respect to the validity of coalescence
models for A--A collisions.
It is then shown that high (i.e. close to unity) baryon/meson
ratio at intermediate $p_{T}$ were already reported in pp
collisions, hence justifying the need for a reference
measurement at LHC energies.
\section{Particle yields and ratios for LHC heavy-ion collisions}
\label{sec:2_bohq06}
%
%
Excitation functions of many particle yields in heavy-ion
collisions are currently available from the Alternating
Gradient Synchrotron energies, through the Super Proton
Synchrotron (SPS) ones, up to a Relativistic Heavy Ion
Collider (RHIC) top energy of $\sqrt{s}=200~\rm{GeV}$~\cite{Letessier:2005qe,Speltz:2005pj,Adams:2005dq,Adcox:2004mh}.
Moreover, centrality dependences were also reported and since
smooth behaviours were observed over more than two orders of
magnitude in energy, clear guidances are given on what
these values might be for the LHC Pb--Pb collisions at
$\sqrt{s}=5.5~\rm{TeV}$.
%

%
%
If some uncertainties remain for the extrapolations of the
total multiplicities and specific absolute yields, the large
success of statistical thermal models at SPS and RHIC should
provide reliable predictions for most particle
ratios~\cite{Becattini:2000jw,Andronic:2005yp}.
In the following paragraph, we discuss observables which, by
a deviation from model predictions, may reveal new physics
related to QGP characteristics and/or evolution.
\subsection{Thermal production and strangeness equilibration}
\label{subsec:22_bohq06}
%
%
Assuming that particle production in heavy ion collisions,
as shown previously at SPS and RHIC~\cite{Becattini:2000jw,Andronic:2005yp}
is described by a thermal source, some statistical
observables can be extrapolated up to LHC energies.
Indeed, a chemical freeze-out temperature and a finite
baryo-chemical potential in the statistical thermal model
approach can be obtained at equilibrium using a constant
average energy per hadron close to $1~\mathrm{GeV}$~\cite{Cleymans:1998fq}
which approximately gives $\mathrm{T}=165~\mathrm{MeV}$ and
$\mu_{\mathrm{B}}=1~\mathrm{MeV}$ at
$\sqrt{s}=5.5~\rm{TeV}$~\cite{Cleymans:2006qe}.
%

%
%
\begin{figure*}
\begin{center}
\resizebox{0.7\textwidth}{!}{%
  \includegraphics{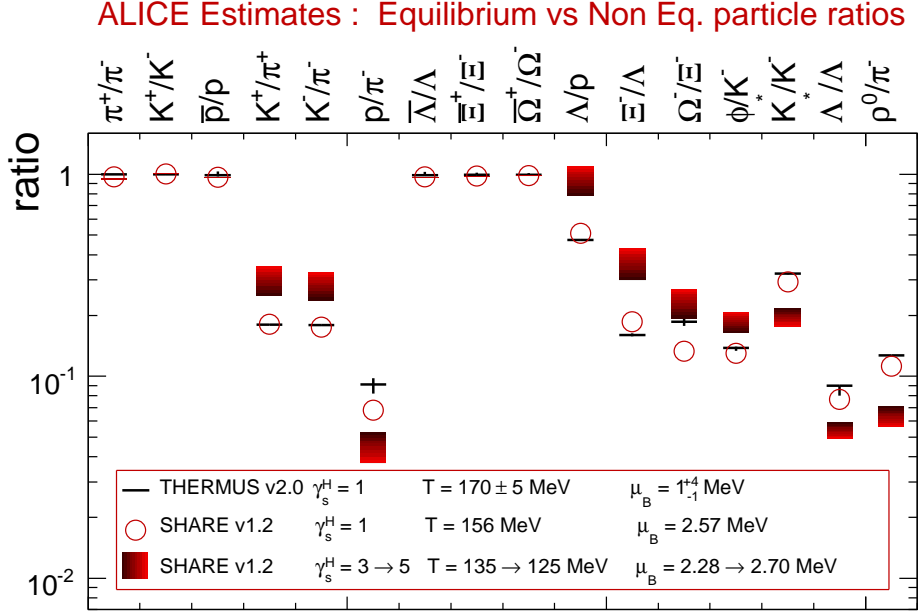}
}
\caption{Estimates of particle ratios in central Pb--Pb
collisions at LHC energies for equilibrium
(THERMUS~\cite{Wheaton:2004qb}) and non equilibrium
(SHARE~\cite{Torrieri:2004zz}) statistical models with
different assumptions related to chemical freeze-out
conditions:
i) for the equilibrium model, a chemical freeze-out
temperature of $\mathrm{T}=170\pm 5~\mathrm{MeV}$ and a
chemical potential
$\mu_{\mathrm{B}}=1^{+4}_{-1}~\mathrm{MeV}$ were extrapolated
from RHIC thermal fits~\cite{Cleymans:2004pp};
ii) for the non equilibrium model, the strangeness
equilibration is reflected in the strange quark phase space
occupancy $\gamma_{s}$ and the image of strangeness
equilibrium in the deconfined source,
$\gamma^{\mathrm{QGP}}_{s}=1$, results in an over-saturation
$\gamma^{\mathrm{H}}_{s}=3 \rightarrow 5$ after sudden
hadronization~\cite{Rafelski:2005jc}.
Strangeness saturation in the hadronic phase i.e.
$\gamma^{\mathrm{H}}_{s}=1$ for the non equilibrium model is
shown as a reference (open circles) and systematics
comparison for the chemical freeze-out conditions
($\mathrm{T}=156~\mathrm{MeV}$ and 
$\mu_{\mathrm{B}} \simeq 2.6~\mathrm{MeV}$).
Expected values for resonances are presented with no
in-medium effects (see text).
}
\label{fig:LhcEquilVsNonEquil_bohq06}
\end{center}
\end{figure*}
%
%
%
For what is related to the overall strange quark production,
most of the information is contained in the Wroblewski
factor~\cite{Wroblewski:1985sz,Anisovich:1974ht}:
$$ 
\lambda_\mathrm{s} \equiv {2\frac{\bigl<\mathrm{s}\bar{\mathrm{s}}\bigr>}
{\bigl<\mathrm{u}\bar{\mathrm{u}}\bigr> +
\bigl<\mathrm{d}\bar{\mathrm{d}}\bigr>}}
$$
From RHIC to LHC energies, this factor should stay almost
constant and reach $\simeq 0.43$ and $\simeq 0.20$ for
respectively Pb--Pb and pp collisions~\cite{Wheaton:2004qb}.
%

%
%
In the same framework, many particle ratios can be estimated
with $\mathrm{T}$ and $\mu_{\mathrm{B}}$ where
it is important to recall that ratios including (heavy)
multi-strange baryons are sensitive to small variations
of the temperature~\cite{Cleymans:2006xj}.
%

%
%
However the models may consider differently the degree of
strangeness equilibration in the final state, which can be
related to the QGP cooling under several assumptions. 
Figure~\ref{fig:LhcEquilVsNonEquil_bohq06} displays the
estimates for particle ratios for two models.
The equilibrium model (THERMUS) shown here assumes complete
equilibrium of strangeness in the final (hadronic) state 
($\gamma^{\mathrm{H}}_{s}$=$1$) whereas for the 
non-equilibrium model (SHARE), equilibration for strange
quarks in the QGP ($\gamma^{\mathrm{QGP}}_{s}$=$1$) can lead 
to an over-saturation after sudden hadronization
($\gamma^{\mathrm{H}}_{s}$=$3 \rightarrow 5$).
These two models are detailed in specific documents for their
implementations and some of their
results~\cite{Wheaton:2004qb,Torrieri:2004zz,Cleymans:2004pp,Rafelski:2005jc}. 
%

%
%
Within errors, THERMUS ratios correspond exactly to the
former values of $\mathrm{T}$ and $\mu_{\mathrm{B}}$.
Moreover, and as expected for a vanishing baryo-chemical
potential, all anti-particle to particle ratios at LHC are
shown to be unity.
This was not yet the case at RHIC energies~\cite{Hippolyte:2003yf}.
One can note that hadronic equilibrium ratios obtained with
SHARE are similar although $\mathrm{T}$ differs
significantly.
%

%
%
For non-identical species ratios, the differences between
THERMUS and SHARE equilibrium results (lines vs. circles:
mainly $\mathrm{p}/\pi^{-}$ and $\Omega^{-}/\Xi^{-}$) can
be interpreted as a systematical variation of the chemical
temperature.
On the contrary non-identical species ratios are very
sensitive to the equilibrium hypothesis: significant
deviations are observed especially where strange baryons
are involved~\cite{Cleymans:2006xj}.
Particle ratios including resonances have to be taken
cautiously.
In-medium effects can be sizeable at the QGP stage (the
lifetime of some resonances are of the order of, or even
smaller than the one of the fireball) or later during the
hadronic stage (e.g. rescattering and
regeneration)~\cite{Adams:2003cc:2004ep:2006yu}:
these effects are not taken into account here.
%

%
%
From these comparisons and assuming that contributions from
hard processes are marginal for these studies, the
distinction between non-equilibrium and equilibrium scenarios
should be within the reach of the ALICE experiment.
Its dedicated particle identification capabilities are stated
in the next section.
\subsection{ALICE particle identification capabilities}
\label{subsec:23_bohq06}
%
%
The design of the ALICE experiment is ideal for particle
identification (PID) in the soft physics sectors 
($p_{T} < 2~\mathrm{GeV}/c$).
Several detectors can be used for PID independently or
simultaneously via, for instance, linear energy loss in gas
detectors or silicon sensors, ring imaging Cherenkov or
time of flight measurements~\cite{Carminati:2004fp}.
Secondary vertex topology identification (for strange
particle e.g. $\mathrm{K}^0_{s}$, $\Lambda$ or charged
multi-strange ones e.g.  $\Xi$ and $\Omega$) has the
significant advantage to allow single PID over a wide range
of $p_{T}$~\cite{Vernet:2005ain}.

%
%
For a statistics of $10^7$ central Pb--Pb collisions at top
LHC energies, the identification ranges for particle spectra
at mid-rapidity were estimated to be 0.2--12, 0.5--11, 1--8
and 1.5--7 $\mathrm{GeV}/c$ respectively for $\mathrm{K}^0_{s}$, $\Lambda$,
$\Xi$ and $\Omega$~\cite{Vernet:2005ain}.
Transverse momentum spectra from pp collisions could be of
similar ranges with a statistics of $10^9$ events in one year
of data taking~\cite{Gaudichet:2005ain}.
\section{Transverse momentum ratios for LHC collisions}
\label{sec:3_bohq06}
In RHIC central heavy ion collisions, quark coalescence was
suggested as a possible hadronization mechanism and explains
qualitatively the mid-rapidity baryon/meson ratio at
intermediate $p_{T}$.
In pp collisions, such a hadronization scenario is not
favoured due to the low phase space density in the final
state.

In the following paragraphs, we will discuss coalescence
mechanisms and predictions for Pb--Pb collisions at LHC.
It will be shown that i) high baryon/meson ratios were
also measured in elementary collisions for energies
respectively 3 and 9 times the top RHIC ones;
ii) a PYTHIA model~\cite{Sjostrand:1993yb:2003wg} hardly
reproduces such a behaviour even with the recent multiple
parton interaction mechanisms and latest parton distribution
functions (with higher gluon density at low $Q^{2}$).   
\subsection{Coalescence and baryon enhancement at intermediate transverse momentum}
\label{subsec:31_bohq06}
%
%
The intermediate $p_{T}$ region can be defined between a soft
domain, reproduced quite accurately by hydrodynamics, and an
upper limit where the fragmentation of energetic partons
dominates hadro-production.
Constituent quark coalescence was suggested
to be a hadronization mode in this region for describing
hadron yields and spectra at RHIC by several models~\cite{Fries:2003kq,Greco:2003mm,Hwa:2002tu} with the possible
inclusion of recombination between bulk and mini-jet quarks.
Being a multi-parton process, recombination is closely
related to the high phase space density available in
heavy-ion collisions but obviously disfavoured in pp
reactions.
%

%
%
Besides the magnitude of the elliptic flow, one major
success of this picture is the description of the
baryon over meson ratios as a function of $p_{T}$.
At RHIC, it was quite surprising to see a kinematical region
where the production of baryons exceeds the one of mesons
with a similar flavour content~\cite{Adams:2006wk,Adler:2003cb}.
From simple (constituent) quark counting considerations, the
interplay of coalescence and fragmentation appears to explain
the magnitude and almost the observed turnover of the
$\bar{\mathrm{p}}/\pi^{-}$ and the
$\Lambda/\mathrm{K}^{0}_{S}$ ratios~\cite{Adams:2006wk}.

For LHC energies, the turnover in the $\mathrm{p}/\pi^{0}$
ratio was predicted to change slightly from
$p_{T} = 3~\mathrm{GeV}/c$ to $\sim 4~\mathrm{GeV}/c$ with
uncertainties associated to the evolution of the transverse
radial flow velocity~\cite{Fries:2003fr}.
\subsection{Baryon over meson ratios in elementary collisions}
\label{subsec:32_bohq06}
%
%
For pp collisions at RHIC ($\sqrt{s}=200~\mathrm{GeV}$), the
$\Lambda/\mathrm{K}^{0}_{S}$ ratio was also measured.
A value of $\sim 0.6$ was obtained which is approximately one
third of the value for central heavy-ion collisions~\cite{Lamont:2004qy}.

\begin{figure*}
\begin{center}
\resizebox{0.7\textwidth}{!}{%
  \includegraphics{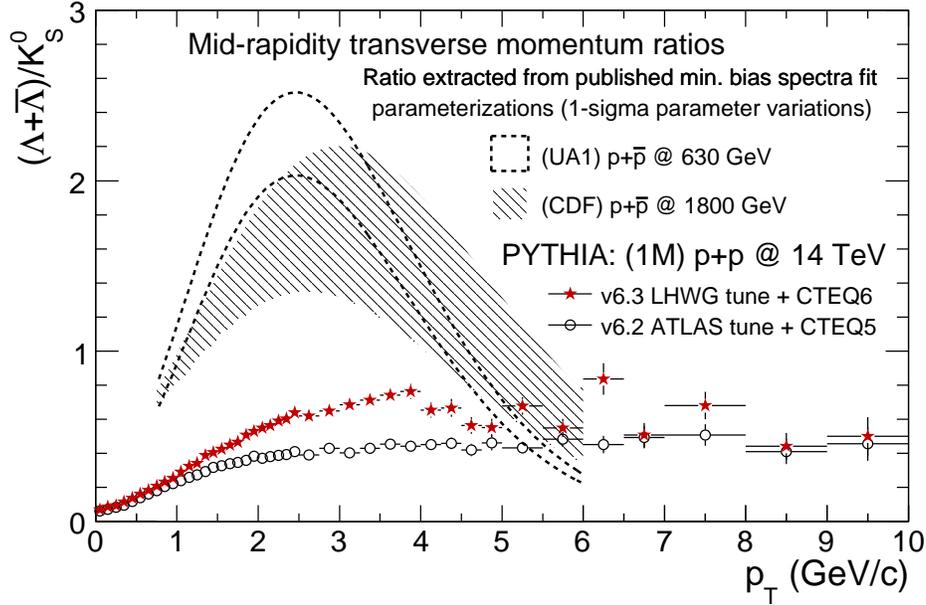}
}
\caption{Baryon over meson transverse momentum ratios for single strange particles $(\mathrm{\Lambda} + \mathrm{\bar{\Lambda}}) / \mathrm{K}^{0}_{s}$ at mid-rapidity for minimum bias elementary collisions. For $\mathrm{p}+\mathrm{\bar{p}}$ results at $\sqrt{s}=630~\mathrm{GeV}$ (UA1 Collaboration)~\cite{Bocquet:1995jq} and $\sqrt{s}=1800~\mathrm{GeV}$ (CDF Collaboration)~\cite{Acosta:2005pk}, the measured ratios are presented as areas where the published fit parameterizations were used for the spectra (allowing 1-$\sigma$ error for each parameter). Ratios extracted from PYTHIA simulations are shown for 1M simulated $\mathrm{p}+\mathrm{p}$ minimum bias events at LHC top energies i.e. $\sqrt{s}=14~\mathrm{TeV}$. Errors are statistical only. Further details and references for these simulations can be found in the text.}
\label{fig:LambdaOverKshortRatio_bohq06}
\end{center}
\end{figure*}

%
%
However, the UA1 and CDF Collaborations measured the
($\Lambda+\bar{\Lambda}$) and $\mathrm{K}^{0}_{S}$ spectra at
mid-rapidity in minimum bias $\mathrm{p}+\bar{\mathrm{p}}$
collisions  at the higher energies of respectively
$\sqrt{s}=630~\mathrm{GeV}$~\cite{Bocquet:1995jq}
and $1800~\mathrm{GeV}$~\cite{Acosta:2005pk}.
If one computes the $(\Lambda+\bar{\Lambda})/\mathrm{K}^{0}_{S}$
ratio from the fit parameterizations they reported, the observed
magnitude is closer to the RHIC heavy-ion one, although
coalescence mechanisms are difficult to invoke here.
These ratios are shown in Fig.~\ref{fig:LambdaOverKshortRatio_bohq06}.
\subsection{PYTHIA simulations and extrapolations}
\label{subsec:33_bohq06}
%
%
Since the coalescence mechanism originates from the
hypothesis of multi-parton processes, it sounds legitimate
to investigate this path for LHC pp collisions at the top
energy of $\sqrt{s}=14~\mathrm{TeV}$.
The PYTHIA model offers such a possibility to include
multiple parton interactions~\cite{Sjostrand:2004ef}.

%
%
The first difficulty is the extrapolation from Tevatron to
LHC energies.
The PYTHIA {\it configuration 1} is proposed by the ATLAS
Collaboration~\cite{Moraes:2005app} and based on the CTEQ5
parton density functions (PDF)~\cite{Pumplin:2002vw}, PYTHIA
version 6.2 and Tevatron Rick Field's tune A~\cite{Field:2002twf}.
It is compared here to the {\it configuration 2}  set by the
``Les Houches Working Group'' (LHWG)~\cite{Buttar:2006zd}. 

%
%
Major modifications were performed from the version 6.2
to 6.3 of PYTHIA.
One of these is the new treatment of multiple parton 
interactions which is supposed to lead to a better
underlying event description~\cite{Sjostrand:2004ef}.
The LHWG based their extrapolation on this latest
implementation and the CTEQ6 PDF which has a higher gluon
density function at low $Q^{2}$ than CTEQ5~\cite{Buttar:2006zd}.
%
%
It is worth mentioning that for these simulations,
anti-particle to particle ratios such as
$\bar{\mathrm{p}}/\mathrm{p}$ and $\bar{\Lambda}/\Lambda$
were checked to be flat in the $p_{T}$ range presented in
Figure~\ref{fig:LambdaOverKshortRatio_bohq06}.
Therefore no hard scattering effects from valence quarks 
are hidden in the addition of the $\Lambda$ and
the $\bar{\Lambda}$ spectra.
This sum was only performed here for having a better
statistics and for consistency with UA1 and CDF data
discussed in section~\ref{subsec:32_bohq06}.
%

%
%
The $(\Lambda+\bar{\Lambda})/\mathrm{K}^{0}_{s}$ ratios obtained
for both configurations and $1~M$ minimum bias pp collisions at
$\sqrt{s}=14~\mathrm{TeV}$ are displayed on 
Figure~\ref{fig:LambdaOverKshortRatio_bohq06} together
with UA1 and CDF data.
For the {\it configuration 1}, the ratio increases up to
$p_{T}=4~\mathrm{GeV}/c$ and then flattens at $\sim 0.4$.
In the {\it configuration 2}, a clear increase is observed,
reaching $\sim 0.7$ at $p_{T}=4~\mathrm{GeV}/c$ and showing
that the cumulated effects of the higher gluon density at
low $Q^{2}$ and the multiple parton interactions give
qualitatively the expected behaviour.
\section{Conclusion}
\label{sec:3_bohq06}
Particle yields and transverse momentum spectra or ratios
including the strange flavor will be part of the first ALICE
results.
These measurements will benefit from the particle
identification capability of the experiment in the soft
sector up to the limit where the fragmentation of energetic
partons dominates hadro-production.
We expect that discrimination between non-equilibrium and
equilibrium scenarios in heavy-ion collisions will be
possible and that a better characterization of baryon
production in pp collisions will help defining the
underlying event structure.
\end{document}